\documentclass[acmsmall,screen]{acmart}

\AtBeginDocument{%
  }

\usepackage{url}            
\usepackage{booktabs}           
\usepackage{nicefrac}       
\usepackage{microtype}      
\usepackage{lipsum}		    
\usepackage{doi}
\usepackage{subcaption}
\usepackage{pifont}
\usepackage{colortbl}
\usepackage{float}
\usepackage{balance}
\usepackage{xspace}
\usepackage{acronym}
\usepackage{xcolor}
\usepackage{adjustbox}
\usepackage{float}
\usepackage{siunitx}
\usepackage{tabularx}
\usepackage{enumitem}
\usepackage[subtle]{savetrees}
\usepackage{hyperref} 
\usepackage{multirow}
\usepackage[table]{xcolor}

\newcommand{\dcircle}[1]{\ding{\numexpr171 + #1}}
\newcommand{\bcircle}[1]{\ding{\numexpr181 + #1}}

\usepackage[strict]{changepage}
\usepackage{framed}

\definecolor{formalshade}{rgb}{0.85,1,0.85} 
\definecolor{darkblue}{rgb}{0.0,0.6,0.30}   

\newenvironment{answer}{%
  \MakeFramed{\advance\hsize-\width\FrameRestore}%
  \noindent\hspace{-4.55pt}
  \begin{adjustwidth}{}{7pt}%
}
{%
  \end{adjustwidth}\endMakeFramed%
}

\begin{document}
\title{Exploring Hidden Geographic Disparities in Android Apps}

\author{Marco Alecci}
\affiliation{%
  \institution{SnT, University of Luxembourg}
  \country{Luxembourg}
}
\email{marco.alecci@uni.lu}

\author{Pedro Jesús Ruiz Jiménez}
\affiliation{%
  \institution{SnT, University of Luxembourg}
  \country{Luxembourg}
}
\email{pedro.ruiz@uni.lu}

\author{Jordan Samhi}
\affiliation{%
  \institution{SnT, University of Luxembourg}
  \country{Luxembourg}
}
\email{jordan.samhi@uni.lu}

\author{Tegawendé F. Bissyandé}
\affiliation{%
  \institution{SnT, University of Luxembourg}
  \country{Luxembourg}
}
\email{tegawende.bissyande@uni.lu}

\author{Jacques Klein}
\affiliation{%
  \institution{SnT, University of Luxembourg}
  \country{Luxembourg}
}
\email{jacques.klein@uni.lu}

\keywords{Android Applications, APK, Software Repository, Metadata}

\begin{abstract}
While mobile app evolution over time has been extensively studied, geographical variation in app behavior remains largely unexplored. This paper presents a large-scale study of location-based Android app differentiation, uncovering two important and previously underexplored phenomena with significant security and fairness implications.
First, we introduce the concept of \textit{GeoTwins}: apps that are functionally similar and share branding, yet are released under different package names (i.e., as two separate apps) across different countries. 
Despite their apparent similarity, GeoTwins often diverge in critical aspects such as requested permissions, third-party libraries, and privacy disclosures.

Second, we investigate the Android App Bundle ecosystem and uncover unexpected regional differences in the supposedly consistent \texttt{base.apk} files, which are generally assumed to be invariant. Contrary to expectations, our analysis shows that even \texttt{base.apk} files can vary by region, revealing hidden customizations that may affect app behavior or security. 

These discrepancies have concrete consequences for researchers, developers, and regulators alike. For instance, as geographically distinct app variants can differ in their \texttt{base.apk} files, the same app may be considered benign in one malware detection study but potentially malicious in another, depending on where it was downloaded. Such hidden regional variation silently undermines the reproducibility of prior research and introduces a geographical bias that can distort conclusions about security, privacy, and app behavior. At the same time, these inconsistencies raise ethical concerns about transparency and user consent: visually identical apps on Google Play, complete with the same branding, screenshots, and descriptions, may actually present subtle differences that users are not aware of. 

To analyze these phenomena, we developed a distributed app collection pipeline spanning multiple regions and analyzed thousands of apps. We also release our dataset of \num{81963} GeoTwins to facilitate further research. Our findings reveal systemic regional disparities in mobile software, with concrete implications for researchers (who must account for geographical variation when designing studies), developers (who must consider regional compliance and user expectations), platform architects (who must address distribution inconsistencies), and policymakers (who must consider digital fairness and global user protection).
\end{abstract}

\maketitle

\section{Introduction}
\label{sec:introduction}

Mobile applications (apps) shape digital experiences across the globe by providing access to communication, commerce, health, entertainment, and more. While extensive research has explored how apps evolve over time ~\cite{cai2020embracing, cai2020longitudinal, pendlebury2019tesseract, mcdonnell2013empirical, wei2012permission, li2021androct}, the orthogonal dimension of how app behavior varies by \textit{location} has received far less attention.
Recent studies have shown that Android apps may be unavailable in certain regions or offer region-specific content or permissions \cite{kumar2022large}. However, these studies primarily focus on app availability and often overlook the structural and functional divergence that can occur beneath the surface at the code implementation level. To address this gap, Guo et al. \cite{guo2025code} conducted a comprehensive study of geo-feature differences in Android apps, revealing widespread variations in security-relevant aspects like advertising, data handling, and authentication across different countries. Their work with FREELENS, a novel framework designed to overcome challenges like code obfuscation, demonstrates that these regional code-level differences can frequently compromise security baselines and introduce disparities in privacy protections.
 
Our paper takes a step further. We address two unexplored aspects of regional differentiation in Android apps: \bcircle{1} We investigate the phenomenon of \textbf{GeoTwins} i.e., apps that are functionally similar, share a brand identity, but are released under different package names across different countries; and
\bcircle{2} We analyze the \textbf{Android App Bundle} ecosystem and reveal unexpected differences in the \texttt{base.apk} files of apps, which are supposed to remain consistent across regions.

\begin{figure}[H]
    \centering
        \begin{subfigure}{0.25\linewidth}
            \centering
            \includegraphics[width=\linewidth]{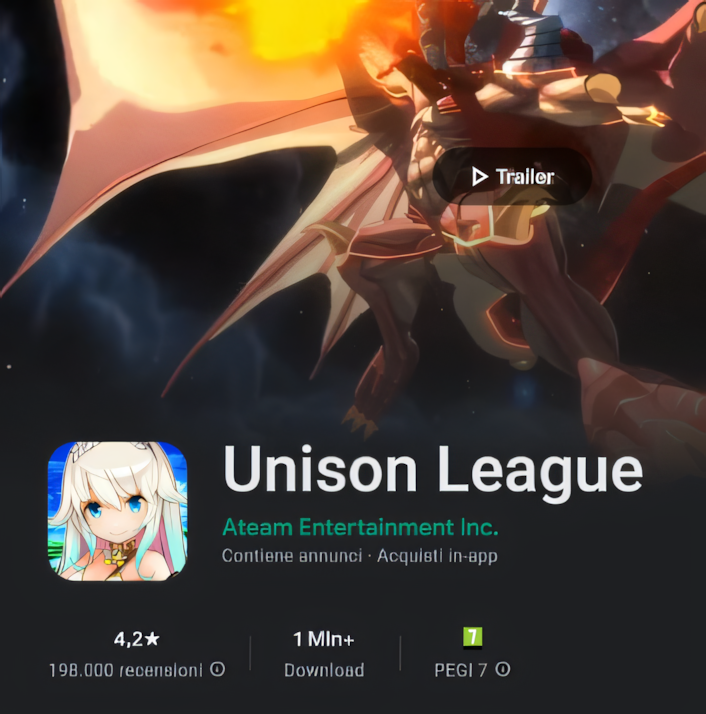}
            \caption{en.co.atm.unison}
            \label{fig:unisonLeagueEN}
        \end{subfigure}
        \begin{subfigure}{0.25\linewidth}
            \centering
            \includegraphics[width=\linewidth]{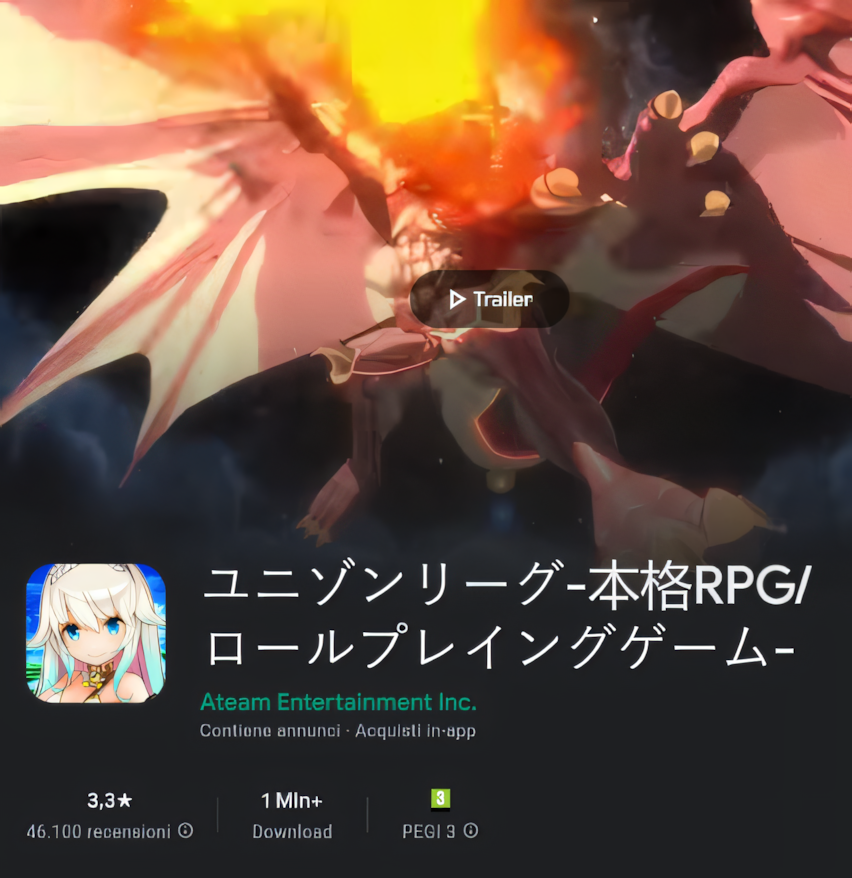}
            \caption{jp.co.atm.unison}
            \label{fig:unisonLeagueJP}
        \end{subfigure}
    \caption{\textit{Unison League} Google Play pages.}
    \label{fig:unisonLeagueScreenshots}
\end{figure}

\noindent
\textbf{GeoTwins: Region-Specific App Variants.}
A central and novel finding of our study is the widespread practice of deploying functionally similar apps under \textit{distinct package names} across countries. We refer to these regional variants as \textit{\textbf{GeoTwins}}. Despite appearing visually similar and sharing a brand identity, GeoTwins often differ in meaningful ways, such as their requested permissions, third-party libraries, and privacy policies. For instance, multinational companies like \textit{McDonald’s} release country-specific versions of their app that differ in privacy disclosures and embedded services. Similarly, mobile games such as the game \textit{Unison League} are published separately for Japanese- and English-speaking audiences under different package names: \texttt{jp.co.atm.unison} (Japan) and \texttt{en.co.atm.unison} (English) as seen in in Figure ~\ref{fig:unisonLeagueScreenshots}.

While this may seem normal in terms of language settings, resources, and images, our analysis revealed that the Japanese version of Unison League includes the \texttt{ACCESS\_FINE\_LOCATION} permission, whereas the international version does not. These cases highlight the subtle yet significant ways that app experiences and protections diverge based on geography. These discrepancies are nontrivial: they raise important concerns about transparency, user consent, and digital inequality. A user in one country may unknowingly receive a more privacy-invasive or less secure version of an app than a user elsewhere despite using what appears to be the same product as seen in Figure ~\ref{fig:unisonLeagueScreenshots}.

Moreover, by looking at Figure~\ref{fig:unisonLeagueScreenshots}, we can observe that even the average review score differs between the two versions. 
While our primary focus is on analyzing in-app differences between regional versions, the concept of GeoTwins provides a valuable framework to also examine metadata such as user reviews, which represents an important direction for future research while still using our concept of GeoTwins. Such studies could provide a more complete understanding of regional variations in app reception and user satisfaction, complementing our findings on in-app disparities and offering valuable insights for developers seeking to optimize user experience across different markets.

\noindent
\textbf{Regional Behavior via App Bundles.}
Additionally, the use of Android App Bundles~\cite{androidappbundle} (a format mandated by Google Play since August 2021) enables developers to tailor app delivery based on users' regions and languages through dynamic delivery. By defining different configuration splits, developers can ensure that only the necessary components are downloaded based on the user's region or language. So while a \texttt{base.apk} file can remain common across all users, language-specific configuration APKs such as \texttt{config-en.apk} for English speakers or \texttt{config-ja.apk} for Japanese speakers can be delivered selectively.  In theory, the \texttt{base.apk} should remain consistent across regions, with regional and language-specific differences handled through configuration splits. In our study, we tried to observe whether any differences existed in the \texttt{base.apk} files, as variations in split APKs are expected due to their configurable nature. Surprisingly, our analysis reveals that even the \texttt{base.apk} files can vary across countries when downloading the ``same" app. These unexpected differences suggest deeper, region-specific customizations that may impact app logic, bundled services, or security practices, raising important questions about consistency and transparency in the app delivery pipeline.

\noindent
\textbf{Our Work.}
We have set up a crawling infrastructure across multiple regions and performed large-scale analysis on thousands of Android apps. In our analysis, we first examine the regional availability of the apps (similar to previous studies~\cite{kumar2022large}), providing further insights on the matter. Then, we go beyond regional availability, structuring the analysis across three levels: \dcircle{1} First, we analyze general trends in location-specific apps (i.e., apps available exclusively in one region) to identify systematic differences between regions --for example, whether Japanese apps consistently differ from European ones--. \dcircle{2} Next, we examine apps with the same package name but downloaded from different locations, focusing on the \texttt{base.apk} component of Android App Bundles to detect region-specific variations introduced by the App Bundle mechanism. \dcircle{3} Finally, we explore the GeoTwins phenomenon by identifying apps that maintain consistent branding and functionality but are distributed under different package names across countries, uncovering significant structural, behavioral, and privacy-related divergences.

\noindent
\textbf{Implications.}
Beyond uncovering these technical phenomena, our findings carry broader implications. They challenge the common assumption that a single app version represents the behavior of that app worldwide. For instance, consider an app $X$ that was analyzed in a malware detection study and classified as benign. If a different version of $X$, downloaded from another country, includes additional permissions or third-party libraries, the same detection tool could potentially classify this regional variant as potentially malicious. Such location-specific divergences can silently bias malware detection, privacy analyses, permission modeling, and other empirical studies that implicitly rely on a single-country view of mobile apps. These differences also pose challenges for reproducibility, user transparency, and regulatory oversight, motivating a broader rethinking of how app ecosystems are studied and governed.
Moreover, these differences also have implications beyond research. Developers can use our dataset to better understand cultural and ecosystem-specific differences, such as how certain libraries (e.g., social media SDKs) are adopted differently across regions, and adapt their apps accordingly. Similarly, regulators may need to consider whether users are adequately informed when visually identical apps behave differently based on their location, and whether clearer disclosures or consistency requirements are warranted. These findings highlight the need for more geographically aware methods in both research and practice, and motivate a rethinking of how mobile ecosystems are studied, regulated, and designed.

\noindent
\textbf{Contributions.}
This paper makes the following contributions:

\begin{itemize}
    \item We conduct a \textbf{large-scale, cross-location analysis} that identifies systemic regional disparities in apps, with important implications for developers, platform designers, and regulators alike.
    \item We introduce the concept of \textbf{GeoTwins} --region-specific variants of the same app released under distinct package names-- and demonstrate their functional and privacy-relevant differences.
    \item We investigate the Android App Bundle ecosystem and uncover \textbf{unexpected regional variations within \texttt{base.apk} files}, which are generally assumed to be uniform across locations.
    \item We release a curated dataset of \num{81963} GeoTwins to facilitate further research in this area.
\end{itemize}

All our data and artifacts are publicly available here:
\begin{center}
\href{https://anonymous.4open.science/r/GeoDroid-777}{https://anonymous.4open.science/r/GeoDroid-777}
\end{center} 
\section{Background}
\label{sec:background}

This section outlines key concepts 
necessary to understand our study.

\noindent
\textbf{Core Terminology.}
We clarify the distinction between an \textit{app}, an \textit{APK file}, and a \textit{package name}:
\begin{itemize}[leftmargin=*]
  \item \textit{App}: An Android application software developed for Android OS, typically distributed via the Google Play Store, third-party markets, or sideloading.
  \item \textit{APK file}: A zip-compressed binary archive that contains all components required to install and run an Android app, e.g., compiled code, resources, metadata, etc. Multiple APKs may exist for different versions of the same app.
  \item \textit{Package name} (\texttt{pkgName}): A unique identifier (e.g., \texttt{com.example.myapp}) for each app, critical for installation, updates, and permissions. Two apps with different package names are treated as completely distinct by the Android operating system, even if they share identical branding or functionality. 
\end{itemize}

\noindent
\textbf{App Bundle Format.}
In modern app distribution, developers must use the \textit{App Bundle} format, which enables more efficient delivery of apps by tailoring APKs to device-specific needs (and therefore location-specific as well) ~\cite{androidappbundle}. This is explicitly requested by the official Android documentation with the following message: ``Important: From August 2021, new apps are required to publish with the Android App Bundle on Google Play."~\cite{androidappbundle}. An app bundle contains a base APK file along with zero or more split APKs.

\begin{itemize}[leftmargin=*]
    \item \textit{Base APK}: The core component of an app that includes the essential code and resources required for the app to function. It should be common across all installations and is always included in the download. Its name is \texttt{base.apk}.
    \item \textit{Split APK}: Optional APKs that handle device-specific configurations such as language, screen density, or architecture. These APKs are delivered selectively to match the user's device.
\end{itemize}

\section{Experimental Setup}
\label{sec:setup}

In this section, we detail the experimental setup used throughout our study.
For our analysis, we rely on two different APK datasets, each constructed using distinct methodologies. We now explain these datasets in detail. The first dataset (\textbf{Global Dataset}) comprises APKs sharing the same package name but obtained through location-specific crawling. The second dataset (\textbf{GeoTwins Dataset}) consists of GeoTwins APKs, i.e., regional variants of the same app with different package names across locations.
Finally, we present our approach to pairwise app analysis, describing the features extracted via static analysis and the similarity scoring system we developed.

\subsection{Global Dataset}
In the literature, it is quite common to rely on the AndroZoo~\cite{androzoo2016} dataset when working with Android apps, due to it being one of the largest collections of Android APKs available for researchers~\cite{alecci2024androzoo}. AndroZoo is reported to source its APKs from either Luxembourg, France, or Canada ~\cite{androzoo2016}. However, the exact location from which each individual APK was downloaded is not disclosed. Therefore, when downloading an APK from AndroZoo, it is not possible to determine whether it was retrieved from Luxembourg, France, or Canada. 

To overcome this limitation and construct a more globally representative dataset, we decided to crawl APKs from the Google Play Store during one year, from seven different locations: Los Angeles, Santiago, Luxembourg, Johannesburg, Tel Aviv, Tokyo, and Sydney.  
These locations were carefully chosen to ensure broad coverage across all inhabited continents, with additional representation of large or diverse regions. For example, North and South America were represented by Los Angeles and Santiago, respectively, while Asia was covered by both Tokyo (Far East) and Tel Aviv (Middle East). We deployed multiple virtual private servers (VPSs) via Vultr.com~\cite{vultr}, a provider known for its extensive global VPS presence. 
Each of these VPSs hosted an independent crawler, which leveraged an unofficial Google Play API~\cite{googleplay-api} to continuously search for and download APKs. To implement our crawlers, we adopted the same strategy used by the official AndroZoo crawlers. Additionally, the crawled APKs were submitted to AndroZoo to contribute to their repository and help consolidate resources for the research community. Over the course of the year, this process yielded a total of \num{73756} unique apps (i.e., unique package names) from across the globe.

\begin{figure}[hbtp]
    \centering
    \begin{subfigure}{0.35\linewidth}
        \centering
        \includegraphics[width=\linewidth]{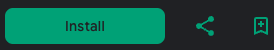}
        \caption{App available}
        \label{fig:install}
    \end{subfigure}
    \begin{subfigure}{0.35\linewidth}
        \centering
        \includegraphics[width=\linewidth]{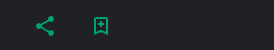}
        \caption{App not available}
        \label{fig:noinstal}
    \end{subfigure}
    \caption{Google Play app page differences: (a) “Install” button present when the app is available in the region, (b) “Install” button absent when the app is not available.}
    \label{fig:installnoinstall}
\end{figure}

Although the APKs were downloaded from these varied locations, this does not necessarily imply that the apps are exclusively available in those countries, as many Google Play apps are globally accessible. To precisely determine the regional availability of each app, we implemented a verification procedure based on a reliable difference in the Google Play Store HTML page. Specifically, when you are logged in Google Play and an app is restricted in a given country, the “Install” button is omitted from its Google Play page. As shown in Figure~\ref{fig:installnoinstall}, the presence of the “Install” button indicates the app is available for download in that region. By detecting this button in the HTML response, we can infer the app’s availability status.

To determine the availability of each app in our dataset, we followed this procedure:

\begin{enumerate}[leftmargin=*]
\item Establish a VPN connection corresponding to each of the seven original download locations.
\item Send a request to the Google Play Store URL for each app:
\url{https://play.google.com/store/apps/details?id=PKG_NAME}
\item Inspect the HTML response for the “Install” button. If present, the app is marked as available in that region; if absent, it is marked as unavailable. A “404” response indicates that the app is no longer listed on Google Play.
\end{enumerate}

This approach enabled us to generate a metadata file annotating each app package with its availability status across the seven surveyed locations.
We then removed apps that were no longer available at the time of the availability check, i.e., those returning a 404 error. The resulting global dataset contains \num{61403} unique apps (i.e., package names) associated with their APK files and comprehensive regional availability metadata.

\subsection{GeoTwins Dataset}
\label{sec:geotwinsdataset}

\begin{figure}[hbtp]
    \centering
    \begin{subfigure}{0.25\linewidth}
        \centering
        \includegraphics[width=0.3\linewidth]{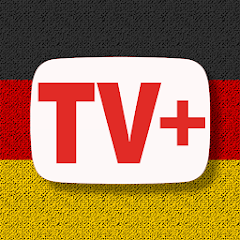}
        \caption{com.cisana.guidatv.de}
        \label{fig:guidatvDE}
    \end{subfigure}
    \begin{subfigure}{0.25\linewidth}
        \centering
        \includegraphics[width=0.3\linewidth]{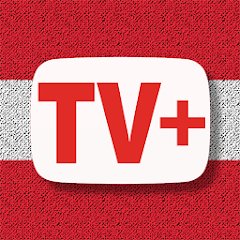}
        \caption{com.cisana.guidatv.at}
        \label{fig:guidatvAT}
    \end{subfigure}
    \caption{\textit{Cisana TV+} app icons associated to different countries.}
    \label{fig:guidatvIcons}
\end{figure}

To collect GeoTwins, we employed a different approach that enabled us to gather a larger volume of apps within a shorter timeframe. While package names can sometimes hint at related apps --such as the \textit{Unison League} game, with \texttt{jp.co.atm.unison} (``jp'' for Japan) and \texttt{en.co.atm.unison} (``en'' for English)-- this alone is insufficient for robust GeoTwins identification. 
To improve accuracy, we decided to incorporate an additional heuristic based on the visual similarity of app icons. Our intuition is that apps released in different countries but developed by the same developer typically share not only similar package names but also identical icons (as seen with \textit{Unison League}) or visually similar icons like those illustrated in Figure~\ref{fig:guidatvIcons}.

\begin{figure}[hbtp]
    \centering
    \begin{subfigure}{0.25\linewidth}
        \centering
        \includegraphics[width=0.3\linewidth]{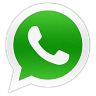}
        \caption{\textit{WhatsApp} icon from an APK found in AndroZoo.}
        \label{fig:guidatvDE}
    \end{subfigure}
    \begin{subfigure}{0.25\linewidth}
        \centering
        \includegraphics[width=0.3\linewidth]{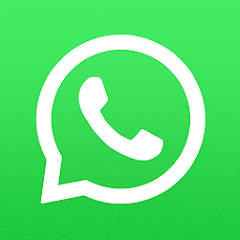}
        \caption{\textit{WhatsApp} icon downloaded from Google Play.}
        \label{fig:guidatvAT}
    \end{subfigure}
    \caption{\textit{WhatsApp} icon from an older version compared to the latest version.}
    \label{fig:updatedIconExample}
\end{figure}

For this step, we were able to rely on AndroZoo --now including the APKs from our ``Global Dataset"-- as we did not need to know the download location to detect GeoTwins through our package names heuristic. Therefore, we started by extracting all distinct package names (\num{8887673} in total) from the AndroZoo dataset.
Our goal was to identify potential GeoTwins within this set. To do this, we used the Google Play Scraper~\cite{googleplay-api} to retrieve the current app icons for each package and computed their hash values using the difference hash (dHash) algorithm~\cite{krawetz2013kindoflikethat}. This algorithm is particularly effective for image comparison, as it captures visual similarities even when icons have undergone minor changes (e.g., resizing, color adjustments, or slight redesigns), such as in the example in Figure~\ref{fig:guidatvIcons}. Notably, we chose not to use icons extracted from the APK files available in AndroZoo since these files often correspond to outdated app versions and may not reflect the apps' current appearance on Google Play. For example, as shown in Figure ~\ref{fig:updatedIconExample}, while older versions of \textit{WhatsApp} found on AndroZoo had a distinct icon, recent updates show a revised design, which highlights the importance of using up-to-date visual data for accurate identification.

As expected, many of the package names retrieved from AndroZoo were no longer available on Google Play as of May 2025 and were therefore excluded. After removing these unavailable apps and computing hashes for the icons of those still accessible, we were left with \num{1138655} package names and their corresponding icon hashes. To improve the reliability of the dataset and focus exclusively on region-specific variants, we applied an additional filter: we retained only those apps whose package names contained a country name or country code. This filtering step reduced the dataset to \num{241059} package names, but it helped ensure that the resulting matches could be confidently associated with different geographical regions.

To identify GeoTwins, we used the normalized Levenshtein distance (NLD) --a variant of the Levenshtein distance~\cite{levenshtein1966binary} obtained by dividing the Levenshtein distance by the longest string length-- for comparing package names and the Hamming distance~\cite{hamming1950error} for comparing icon hashes. We had to rely on NLD in addition to differences in country names or codes, due to cases like the GeoTwin \texttt{sk.martinus.knihovratok} and \texttt{cz.martinus.knihovratek}. These package names present slight differences from language variations --``knihovrátok" in Slovak and ``knihovrátek" in Czech-- both meaning ``bookworm".
Given the vast number of possible pairwise comparisons (i.e., \num{29054600211} pairs for the \num{241059} packages available) we employed a Nearest Neighbors approach~\cite{cover1967nearest} to significantly reduce computation time. For icon similarity, we set a Hamming distance threshold of \num{10}, consistent with prior studies~\cite{zheng2021novel, zhang2023method}. For string similarity, we adopted a normalized Levenshtein distance threshold of \num{0.2}. While the optimal threshold can vary by context and would ideally be determined through extensive manual analysis (an effort beyond the scope of this study) we drew guidance from previous research~\cite{yamaguchi2012discriminative, soualmia2012matching}, which recommends values between \num{0.125} and \num{0.3}. We selected \num{0.2} as a balanced choice: low enough to reduce false positives, yet permissive enough to capture the naming variations commonly associated with region-specific versions. 
Finally, we retained only those pairs with distinct country names or codes to further ensure that the apps targeted different regions. Although this filtering step significantly reduced the number of potential matches, it was a necessary trade-off to produce a high-confidence dataset. The resulting GeoTwins dataset contains \num{81963} GeoTwins i.e., is \num{81963} pairs of regional variants of the same app, released under different package names.

\noindent
\textbf{GeoFamilies.}
Once the GeoTwins dataset was created, we observed that many GeoTwins could actually be grouped into larger collections of related apps. These collections, which we refer to as \textit{GeoFamilies}, consist of multiple regional variants of the same underlying app. Each variant is typically targeted to a different country and published under a slightly different package name, often with minor localization adjustments. For example, as shown in Table~\ref{tab:exampleGeoFamilies} the GeoFamily for \textit{Unison League} includes \texttt{jp.co.atm.unison} and \texttt{en.co.atm.unison}, while the \textit{Papa John’s} app includes variants such as \texttt{cl.com.papajohns}, \texttt{pa.com.papajohns}, and \texttt{cr.com.papajohns}. A larger example is the \textit{American Express} app family, which spans nine different regional variants.

\begin{table}[hbtp!]
    \centering
    \caption{Examples of GeoFamilies.}
    \label{tab:exampleGeoFamilies}
    \begin{adjustbox}{width=0.5\linewidth}
    \begin{tabular}{c|l}
    \textbf{GeoFamily}                                             & \multicolumn{1}{c}{\textbf{Package Names}} \\ \hline
    \multirow{2}{*}{\textbf{Unison League}} & en.co.atm.unison                           \\
                                            & jp.co.atm.unison                           \\ \hline
    \multirow{3}{*}{\textbf{Papa John’s}} & cr.com.papajohns                           \\
                                            & cl.com.papajohns                           \\
                                            & pa.com.papajohns                           \\ \hline
    \multirow{9}{*}{\textbf{American Express}} & com.americanexpress.android.acctsvcs.us    \\
                                            & com.americanexpress.android.acctsvcs.be    \\
                                            & com.americanexpress.android.acctsvcs.au    \\
                                            & com.americanexpress.android.acctsvcs.uk    \\
                                            & com.americanexpress.android.acctsvcs.nl    \\
                                            & com.americanexpress.android.acctsvcs.de    \\
                                            & com.americanexpress.android.acctsvcs.ca    \\
                                            & com.americanexpress.android.acctsvcs.fr    \\
                                            & com.americanexpress.android.acctsvcs.japan
    \end{tabular}
    \end{adjustbox}
\end{table}

The GeoTwins dataset we built contains a total of \num{81963} GeoTwins. These pairs can be grouped into \num{859} distinct GeoFamilies with a median number of 2 distinct package names per GeoFamily and an average of 8.82 package names. It is noteworthy to highlight that within a GeoFamily of $n$ apps, all unique pairs of apps (i.e., combinations of two apps) are considered GeoTwins using the following formula:
\[
\text{\#GeoTwins} = \binom{n}{2} = \frac{n \times (n - 1)}{2}
\]
For example, a family with three apps $a$, $b$, and $c$ yields three pairs of GeoTwins: (a, b), (a, c), and (b, c), as in the case of \texttt{com.papajohns}.
While GeoFamilies offer a broader view of app release strategies across regions, this paper focuses exclusively on GeoTwins, i.e., individual pairs of region-specific variants. To avoid redundancy and ensure statistical independence, we consider at most one GeoTwins pair per GeoFamily in our analysis (e.g., in RQ4). A comprehensive study of full GeoFamilies --including their internal structure, evolution over time, and localization strategies-- is left for future work.

\subsection{Apps Analysis}
\label{sec:analysis}

Our analysis focuses on pairwise comparisons between APKs. Specifically, we compare APKs with the same package name but downloaded from different locations (Global Dataset), as well as pairs of GeoTwins APKs (GeoTwins Dataset). Due to the large volume of apps and the associated time and resource constraints, we decided to rely on static analysis techniques. To perform these comparisons, we developed a custom similarity framework in Python, built on top of ApkTool\cite{apktool} and Androguard\cite{androguard}. Our tool extracts a comprehensive set of features from each APK, including: the list of permissions requested; the names of app components (activities, services, receivers, providers); the APK certificate; third-party libraries used; native libraries; hard-coded URLs embedded within the APK; and all files contained in the APK, with particular emphasis on the ``.smali'' files. Smali files are the human-readable representation of Dalvik bytecode, the intermediate format executed by the Android runtime. We selected these features because they collectively capture both high-level app behavior (e.g., permissions, components, libraries) and low-level code differences (e.g., smali code, embedded files). This broad set of features allows us to detect both functional and structural variations that may arise due to regional customizations or divergent development practices. 

Although existing tools such as SimiDroid~\cite{li2017simidroid} have been developed to compute the similarity between Android APKs, we implemented our own framework to gain fine-grained control over the comparison process. Our tool supports features that SimiDroid does not analyze --such as third-party libraries, hard-coded URLs, and others--. Similar to SimiDroid, we compute a similarity score for each extracted feature, as well as an overall similarity score that aggregates feature-level comparisons. Each score ranges from 0 (completely dissimilar) to 1 (identical). Our similarity tool is publicly available in our repository and can be used to compare any pair of APKs, representing an additional contribution to this work.

\noindent
\textbf{Similarity Score.}
The scoring methods vary by feature type:
\begin{itemize}[leftmargin=*]
    \item \textbf{Set-based features (Permissions, Component Names, Native Libraries, Third-Party Libraries, URLs):} Given the sets of elements in Apk 1 ($A_1$) and Apk 2 ($A_2$), the similarity score $J(A_1,A_2)$ is calculated using the Jaccard Score~\cite{jaccard1901etude}:
    $$J(A_1, A_2) = \frac{|A_1 \cap A_2|}{|A_1 \cup A_2|}$$
    This metric quantifies the proportion of shared elements relative to the total number of distinct elements across both apps.

    \item \textbf{Files (and Smali Files):}
    We compute two separate similarity scores: one for all files in the APK (e.g., assets, resources, images), and another specifically for \texttt{.smali} files, which represent Dalvik bytecode and capture the app’s logic. The scoring method is the same for both. Instead of simply comparing file names and treating them as set-based features, we also analyze the content using a hash-based comparison. Files with matching names and hashes are considered \textit{identical}, while those with matching names but differing hashes are considered \textit{similar}. We then compute a modified Jaccard score, weighting identical files as 1.0 and similar files as 0.5:
    \[
    J_{\text{mod}} = \frac{|F_{\text{identical}}| + 0.5 \cdot |F_{\text{similar}}|}{|F_1 \cup F_2|}
    \]
    where $F_{\text{identical}}$ is the set of identical files, $F_{\text{similar}}$ the set of similar files, and $F_1$, $F_2$ the sets of all files in Apk 1 and Apk 2, respectively. Rather than attempting to quantify how different the files are, we assign a fixed partial score (0.5) to similar files. This approach allows us to efficiently detect file-level changes in the whole APK without performing expensive fine-grained comparisons.
    \item \textbf{JSON Nested Object (APK Certificate):}
    To assess the similarity between the APK certificates, we first flatten their JSON structure. The similarity score is then computed as the fraction of keys with matching values across both objects:
    \[
    S = \frac{\text{number of matching key-value pairs}}{\text{total number of unique keys}}
    \]
\end{itemize}

The overall similarity score is computed as the average of all individual feature scores. Our framework returns both the overall score and the individual feature-level scores to facilitate detailed analysis. All implementation details and the source code of our similarity framework are publicly available in our repository to support reproducibility and further research.

\section{Research Questions}
Our study is guided by four research questions that examine both the availability and internal composition of Android apps across different geographical regions.

\begin{itemize}
    \item\textbf{[RQ1] Regional Availability:} How does the availability of Android apps vary across different locations?
    \item \textbf{[RQ2] General Trends Across Regions:} Are there systematic differences in Android apps across geographical regions (e.g., Japanese apps vs. European apps)?
    \item \textbf{[RQ3] Location-Dependent Differences in Base APKs:} For apps sharing the same \texttt{packageName}, do the \texttt{base.apk} files delivered by the Play Store differ based on the user’s geographical location? If so, what are the nature and implications of these differences?
    \item \textbf{[RQ4] GeoTwins Divergence:} Do GeoTwins exhibit differences in structure, security practices, or other behaviors? If so, what are these key differences?
\end{itemize}


\subsection{[RQ1] Regional Availability}
In this research question, we examine the regional availability of apps within the Global Dataset introduced in Section~\ref{sec:setup}. Using the metadata collected during our availability checks, we assess how app accessibility varies across the seven selected geographical locations. Table~\ref{tab:rq1_numLocations} presents a quantitative summary of app availability. Our dataset comprises \num{61403} unique package names, each verified for presence in the Google Play Store across all seven regions.

\begin{table}[hbtp!]
    \centering
    \caption{App availability across the seven inspected locations.}
    \label{tab:rq1_numLocations}
    \begin{adjustbox}{width=0.4\linewidth}
    \begin{tabular}{c|cc}
    \textbf{\begin{tabular}[c]{@{}c@{}}Number of \\ Locations\end{tabular}} & \textbf{Number of Apps} & \textbf{Percentage} \\ \hline
    \textbf{7}       & \num{48178} & 78.46\% \\
    \textbf{6}       & \num{779}   & 1.27\%  \\
    \textbf{5}       & \num{179}   & 0.29\%  \\
    \textbf{4}       & \num{198}   & 0.32\%  \\
    \textbf{3}       & \num{370}   & 0.60\%  \\
    \textbf{2}       & \num{1007}  & 1.64\%  \\
    \textbf{1}       & \num{10685} & 17.40\% \\
    \end{tabular}
    \end{adjustbox}
\end{table}

As shown, the majority of apps --specifically \num{48178} (78.46\%)-- are available in all surveyed locations, suggesting strong global distribution across much of the mobile app market. In contrast, \num{10685} apps (17.40\%) are exclusive to only one of the seven locations. This highlights a significant subset of regionally restricted apps, indicating the influence of location- or culture-specific targeting, regulatory constraints, or localization strategies.
It is important to note that when we refer to apps available in only ``one location", we mean strictly within the seven locations included in our analysis. For example, an app available ``only" in Tokyo (in the context of our study) may also be accessible in other countries not included in our dataset (e.g., South Korea). Nonetheless, this methodology provides meaningful comparative insights into the geographical availability of Android apps across a diverse and representative set of global locations.

\begin{figure*}[!htbp]
    \centering
    \begin{adjustbox}{valign=c, width=\linewidth}
    \includegraphics{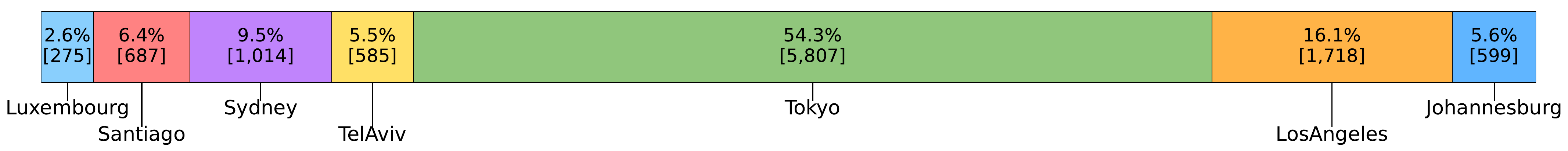}
    \end{adjustbox}
    \caption{Distribution of apps available exclusively in one location.}
    \label{fig:rq1uniquePkgNamePerLocation}
\end{figure*}

To better understand the characteristics of the \num{10685} apps available in only a single location, we analyzed their distribution across the seven locations analyzed. Figure~\ref{fig:rq1uniquePkgNamePerLocation} shows the number of apps exclusive to each location. Tokyo stands out, accounting for \num{5807} exclusive apps --far more than the next highest, Los Angeles, with \num{1718}--. This suggests that Tokyo hosts a particularly distinct app market. Contributing factors may likely include Japan’s unique cultural context, which shapes user expectations around app features, UI/UX design, and content. Developers may thus create apps specifically for the Japanese market or publish heavily localized versions of global apps. Additionally, language complexity and high user standards may incentivize maintaining separate versions.

In summary, while most apps are broadly accessible, a significant minority are regionally exclusive --particularly in Tokyo-- reflecting market dynamics driven by cultural and linguistic factors. Unlike prior work such as Kumar et al., ~\cite{kumar2022large}, which explored the reasons for regional restrictions, our focus is on quantifying and characterizing availability patterns across diverse regions. Investigating the causes of unavailability is beyond our current scope, as it has been addressed in their study.

\begin{answer}
    \textbf{Answer to RQ1:} \textit{App availability is largely global, with 78.46\% of apps accessible in all seven locations studied. However, 17.40\% of apps appear only in a single location --most notably in Tokyo-- indicating a significant degree of regional exclusivity, which could be influenced by differing cultural environments and user preferences.}
\end{answer}


\subsection{[RQ2] General Trends Across Regions}

In this research question, we investigate whether systematic differences exist among Android apps across geographical regions by analyzing how the \num{10685} apps exclusive to a single region (identified in RQ1 and shown in Figure~\ref{fig:rq1uniquePkgNamePerLocation}) differ. Each region’s apps are analyzed independently --without cross-location matching-- allowing us to uncover high-level regional trends in design and behavior. This broad analysis serves as a foundation for the more detailed comparisons explored later in RQ3 and RQ4. While many technical and behavioral features could be examined, we focus on three key aspects: \bcircle{1} Privacy Policies, \bcircle{2} Third-Party Library Usage, and \bcircle{3} Dangerous Permission requests. These were chosen for their clear relevance to privacy and app behavior, and to serve as illustrative examples of region-specific disparities.

\begin{figure}[H]
    \centering
    \begin{adjustbox}{valign=c, width=0.5\linewidth}
        \includegraphics{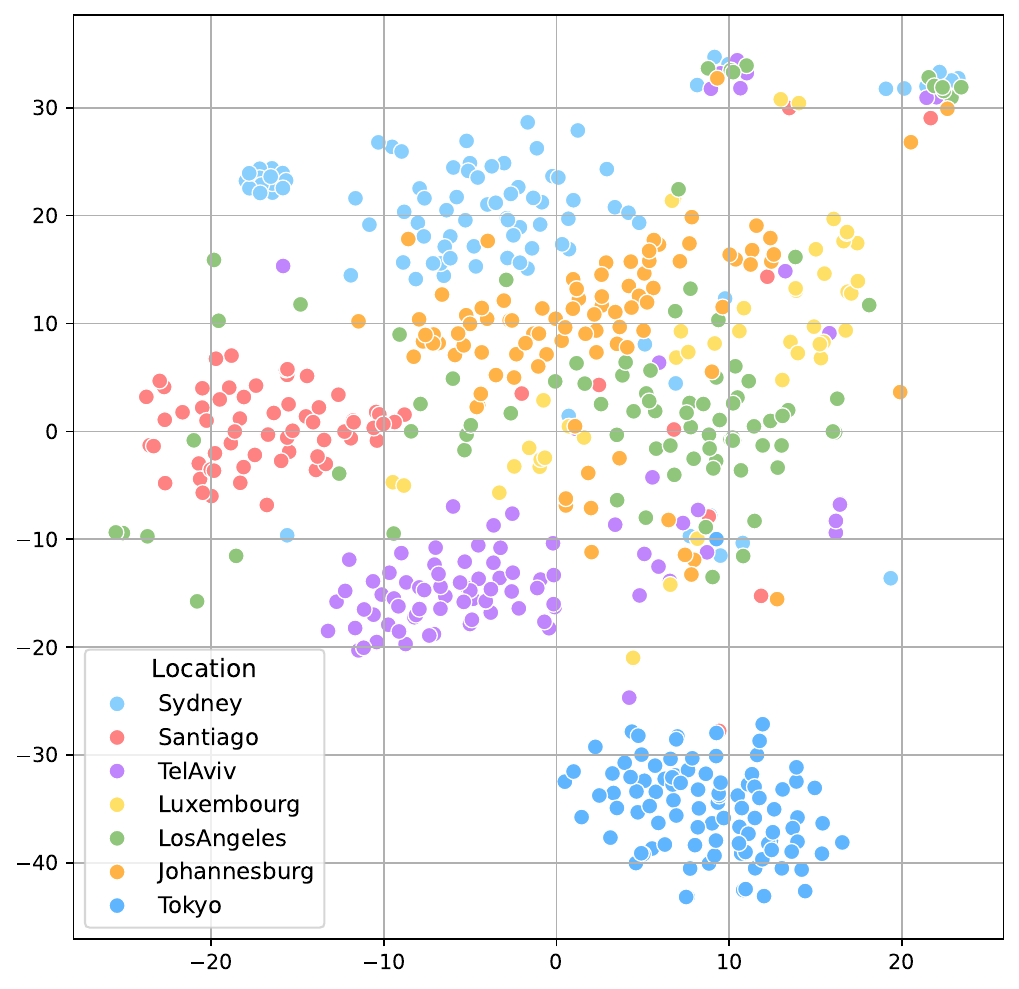}
    \end{adjustbox}
    \caption{2D visualization of privacy policy text embeddings across the seven selected locations.}
    \label{fig:rq2privacyPolicies}
\end{figure}

\noindent
\textbf{\bcircle{1} Privacy Policies.}  
We begin by examining the content of privacy policies, as apps published on platforms like Google Play are typically expected to provide such policies to inform users about their data practices. We collected privacy policy URLs from the Google Play pages of location-exclusive apps. Due to the imbalance in app counts across regions (e.g., nearly \num{6000} in Tokyo vs. \num{275} in Luxembourg), we used a controlled sampling strategy, selecting the top 100 most downloaded apps per region within our dataset, provided they had a valid privacy policy URL. This ensures focus on popular apps representative of typical user experience. We then translated all non-English policies to English using the DeepL Pro API~\cite{deeplpro}, embedded them using OpenAI’s \texttt{text-embedding-3-small} model~\cite{textembedding3small}, and visualized them via t-SNE~\cite{van2008visualizing} (a dimensionality-reduction technique). Figure~\ref{fig:rq2privacyPolicies} shows the resulting 2D scatterplot.

The plot reveals location-specific clustering of privacy policy content --for instance, policies from Tokyo or Tel Aviv form distinct groups-- suggesting these documents reflect regional norms, expectations, and regulatory environments rather than adhering to a uniform global standard.

\begin{figure}[!htbp]
    \centering
    \begin{adjustbox}{width=0.6\linewidth}
        \includegraphics{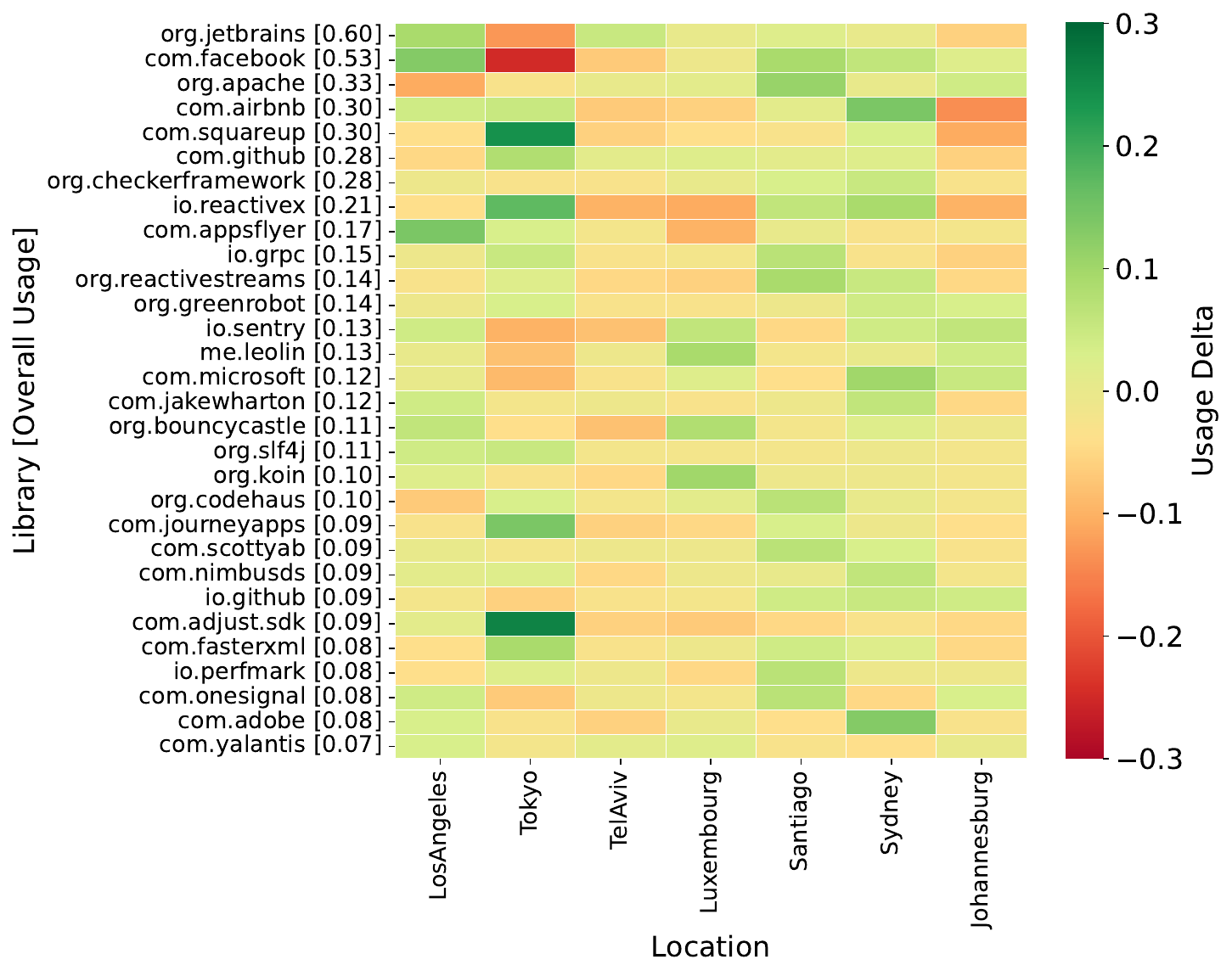}
    \end{adjustbox}
    \caption{Location-based deviations in the usage frequency of the top 30 third-party libraries.}
    \label{fig:rq2libs}
\end{figure}

\noindent
\textbf{\bcircle{2} Third-Party Library Usage.}  
Third-party libraries are widely used to implement functionality such as ads, analytics, networking, and UI. However, they also pose risks related to tracking, data sharing, and compliance. Using again a sample of 100 apps per region, we extracted the list of third-party libraries used. We then computed normalized usage frequencies per region and compared them to global averages. Figure~\ref{fig:rq2libs} presents a heatmap showing location-based deviations for the 30 most common libraries.

Several patterns emerge. For example, \texttt{com.facebook} is more prevalent in Los Angeles and less so in Tokyo --likely reflecting differences in integration with Western versus Eastern platforms--. These differences reflect variations in development practices, platform ecosystems, and privacy implications across locations.

\begin{figure}[!htbp]
    \centering
    \includegraphics[width=0.55\linewidth]{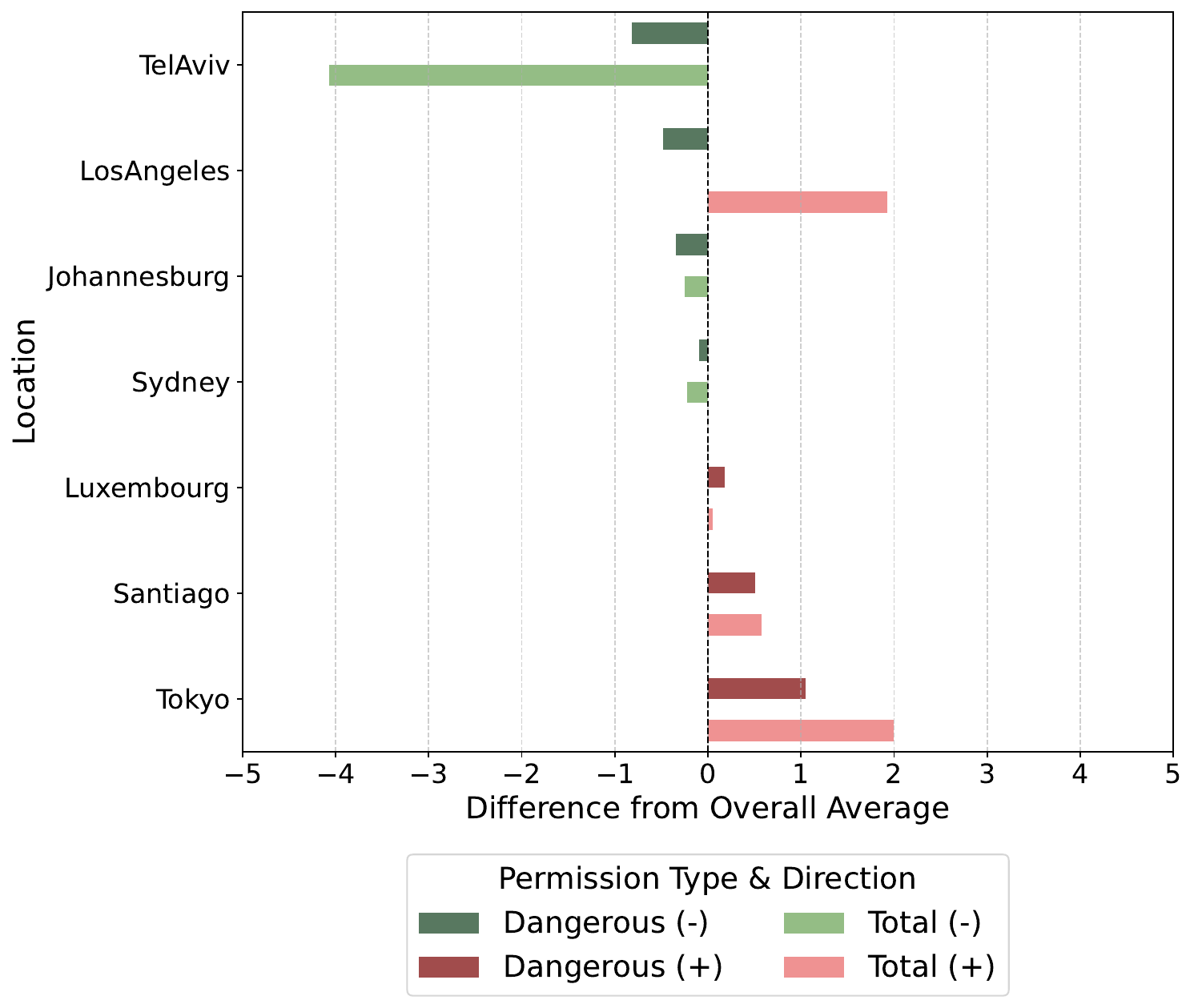}
    \caption{Average deviation from global average in requested (dangerous) permissions, by location.}
    \label{fig:rq2permissions}
\end{figure}

\noindent
\textbf{\bcircle{3} Dangerous Permissions.}  
We also analyzed the use of Android permissions, focusing on both total permissions and those deemed "dangerous" by the Android documentation\cite{androidManifestPermission} --i.e., those granting access to sensitive data or control over critical device features--. Using again the 100-app-per-location sample, we extracted requested permissions as described in Section~\ref{sec:analysis}. We calculated the average number of permissions (dangerous and total) for each location and compared these values to the global average (across all locations). The results are shown in Figure~\ref{fig:rq2permissions}.

As shown, Tokyo apps request more total and dangerous permissions than the global average, while apps from Tel Aviv request fewer. Other locations show milder deviations. These differences suggest that regional development styles or regulatory factors influence how apps engage with the Android permission model as well.

\begin{answer}
\textbf{Answer to RQ2:} \textit{Apps across regions differ not only in availability (as seen in RQ1) but also in key structural and privacy-related aspects. Privacy policies cluster by location, library usage varies, and permission requests show differing privacy exposures --highlighting the systemic influence of geography on Android app behavior--.}
\end{answer}


\subsection{[RQ3] Location-Dependent Differences in Base APKs}
This research question investigates whether apps with the same package name but downloaded from different regions --i.e., the apps from our \textbf{Global Dataset}-- exhibit differences in their \texttt{base.apk} files. 
As it would be extremely time- and resource-consuming to analyze all the apps in our Global Dataset available in all seven locations (\num{48178} apps as presented in RQ1), we decided to analyze a statistically significant random sample of apps. Using a 95\% confidence level with a 5\% margin of error, we computed a sample size of \num{382} apps (i.e. 764 APKs). 
While we initially collected apps during the dataset construction, we had to ensure that the versions compared were the latest ones and that downloads occurred within a narrow time window, reducing the likelihood of version drift. We, therefore, re-crawled all sampled apps across all seven locations simultaneously. This approach is similar to the methodology employed by Kumar et al.~\cite{kumar2022large} for collecting the applications used in their study.
For each app, we downloaded the full app bundle --including the \texttt{base.apk} and split APKs--, but then limited our analysis to the \texttt{base.apk} only. This allowed us to focus our comparison specifically on the \texttt{base.apk} component, avoiding the expected and acceptable differences found in configuration-specific split APKs (as described in Section~\ref{sec:background}). In doing so, our study addresses a key limitation in prior work~\cite{kumar2022large, guo2025code}, which did not distinguish between base and split APKs. This oversight in earlier studies may have led to false positives by attributing configuration-driven differences --intentionally introduced by Google Play-- as evidence of behavioral variation. 
Using the custom similarity score tool introduced in Section~\ref{sec:analysis}, we conducted all 21 region pairwise comparisons for each app. For example, an app from Luxembourg was compared against its versions from the six other locations.

\begin{figure}[h]
    \centering
    \includegraphics[width=0.6\linewidth]{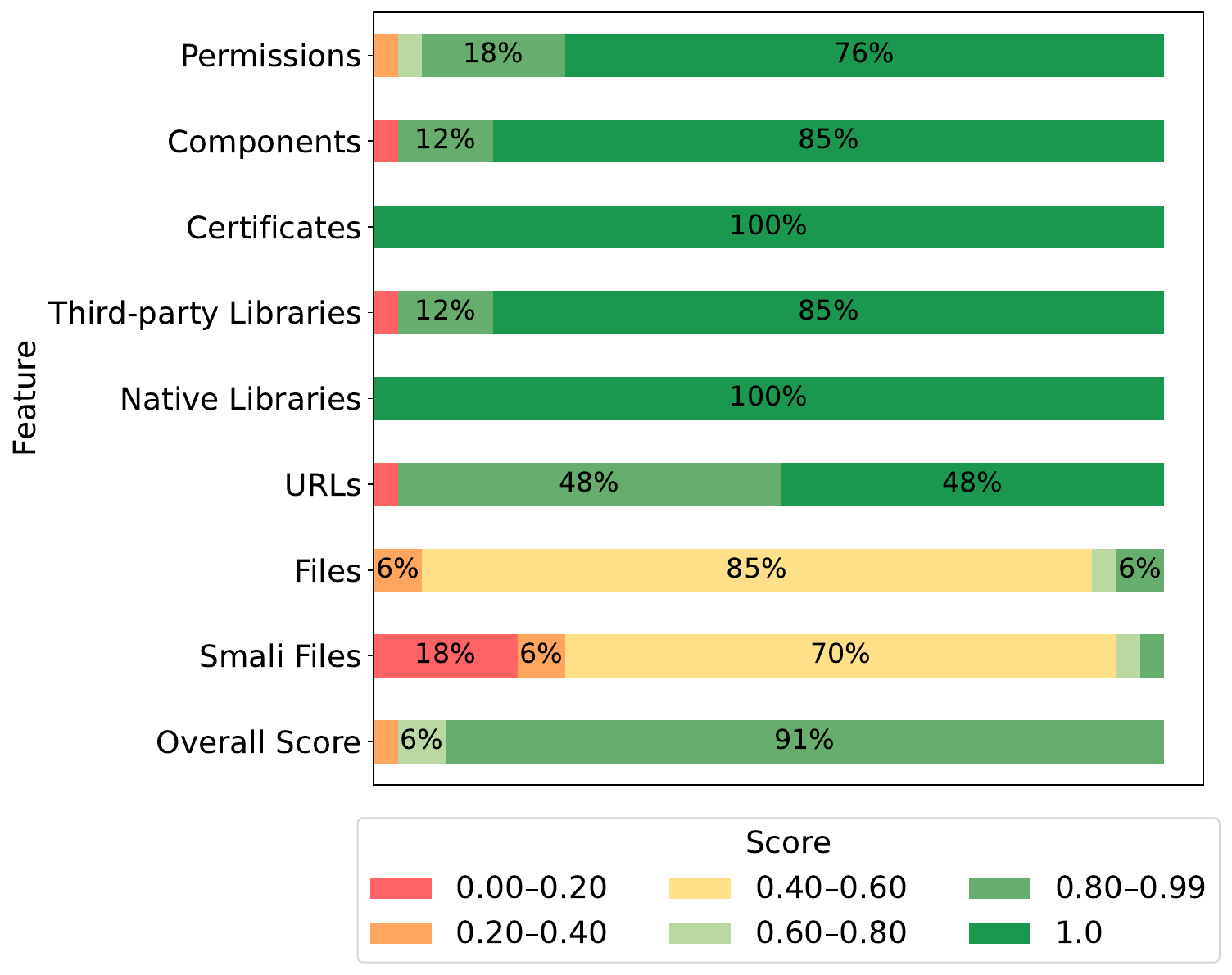}
    \caption{Distribution of similarity scores across features. Scores below 3\% are omitted for readability.}
    \label{fig:rq3distribution}
\end{figure}

\noindent
\textbf{Results.}
Among the \num{382} sampled apps, 33 ($\sim$8.6\%) showed differences in their \texttt{base.apk} files. While small in proportion, this is notable given that \texttt{base.apk} files are expected to be consistent across locations. We assessed these differences by computing similarity scores across their features (Table~\ref{tab:rq3results}) and visualizing the score distributions (Figure~\ref{fig:rq3distribution}).

\begin{table}[h]
    \centering
    \caption{Similarity metrics between GeoTwins. N.A. indicates uniformity across all samples.}
    \label{tab:rq3results}
    \begin{adjustbox}{width=0.6\linewidth}
\begin{tabular}{@{}l|c|c|c@{}}
\textbf{Feature} & \textbf{Average} & \textbf{Median} & \textbf{Correlation with Overall} \\ \midrule
Permissions         & 0.96 & 1.00 & 0.91 \\
Components          & 0.97 & 1.00 & 0.92 \\
Certificates        & 1.00 & 1.00 & N.A. \\
Third-party Libs    & 0.96 & 1.00 & 0.91 \\
Native Libraries    & 1.00 & 1.00 & N.A. \\
URLs                & 0.93 & 0.99 & 0.89 \\
Files               & 0.52 & 0.50 & 0.58 \\
Smali Files         & 0.43 & 0.50 & 0.57 \\ \midrule
\textbf{Overall}    & \textbf{0.84} & \textbf{0.87} & \textbf{1.00}
\end{tabular}
    \end{adjustbox}
\end{table}

As shown, permissions, components, and third-party libraries are nearly identical across regions (avg/median 0.96–1.00), suggesting consistency. URLs are slightly more variable (avg 0.93), possibly due to region-specific endpoints. Files and Smali code exhibit the greatest variation (avg $\sim$0.43–0.52), which largely accounts for overall discrepancies. Notably, 91\% of overall similarity scores fall between 0.80–0.99, aligning with an average of 0.84 and a median of 0.87. 

\noindent
\textbf{Illustrative examples.}
Despite high similarity scores overall, some apps reveal substantial differences. For example, \texttt{club.eatery.prostopizza} shows a permission similarity of 0.74 between the Santiago and Luxembourg versions. The Luxembourg variant requests 26 permissions, including \texttt{USE\_BIOMETRIC}, \texttt{USE\_FINGERPRINT}, \texttt{READ\_GSERVICES}, and \texttt{ACCESS\_ADSERVICE\_TOPICS}, while the Santiago version requests 25, with unique permissions such as \texttt{ACCESS\_BACKGROUND\_LOCATION} and \texttt{READ\_MEDIA\_IMAGES}. This highlights discrepancies in biometrics, location, and media access.

Another case, \texttt{com.shipbuild.buildwishipgb.ckxm}, shows differences in third-party libraries between its Los Angeles and Tel Aviv versions. The LA version includes 12 libraries, while the Tel Aviv one includes 11. Both share common libraries such as \texttt{com.bytedance}, \texttt{com.squareup}, and \texttt{com.facebook}, but the LA variant uniquely includes \texttt{com.fyber}, suggesting region-specific advertising configurations.

\begin{answer}
\textbf{Answer to RQ3:} \textit{While most apps with the same package name have highly similar \texttt{base.apk} files across regions, around 8.6\% exhibit non-trivial differences --particularly in files, code, and occasionally permissions or libraries--. These findings challenge the assumption of a globally consistent \texttt{base.apk} and suggest hidden region-specific customizations in some apps.}
\end{answer}


\subsection{[RQ4] GeoTwins Divergence}
The goal of this research question is to analyze the differences between GeoTwins and assess the extent of their divergence by examining the apps included in the GeoTwins Dataset, introduced earlier in Section~\ref{sec:setup}.
From our full GeoTwins dataset of \num{81963} pairs, we selected a statistically significant sample using the same methodology as in RQ3. Based on a 95\% confidence level and a 5\% margin of error, this resulted in a sample of \num{383} GeoTwins (i.e., \num{766} APKs). To avoid sampling bias from large, overrepresented GeoFamilies (see Section~\ref{sec:geotwinsdataset}), we sampled at most one GeoTwin pair per GeoFamily. This approach ensured both statistical validity and broad coverage across diverse regional variants.
Since the GeoTwins Dataset was derived from AndroZoo metadata, we could have used AndroZoo to retrieve the APK files. Nevertheless, we chose to download the APKs ourselves to ensure we obtained their latest versions and to avoid inconsistencies caused by potential updates since the time AndroZoo crawled them (AndroZoo does not publish the exact crawl date). This also allowed us to extract and analyze both the \texttt{base.apk} and split APKs, though our analysis here --like in RQ3-- focuses on the \texttt{base.apk}.
By collecting and analyzing GeoTwins, this work addresses a key limitation in Guo et al.~\cite{guo2025code}, who compared apps with shared package names across multiple locations (similar to our approach in RQ3) but lacked the ability to compare what they referred to as ``regional variants". 

\begin{figure}[h]
    \centering
    \includegraphics[width=0.6\linewidth]{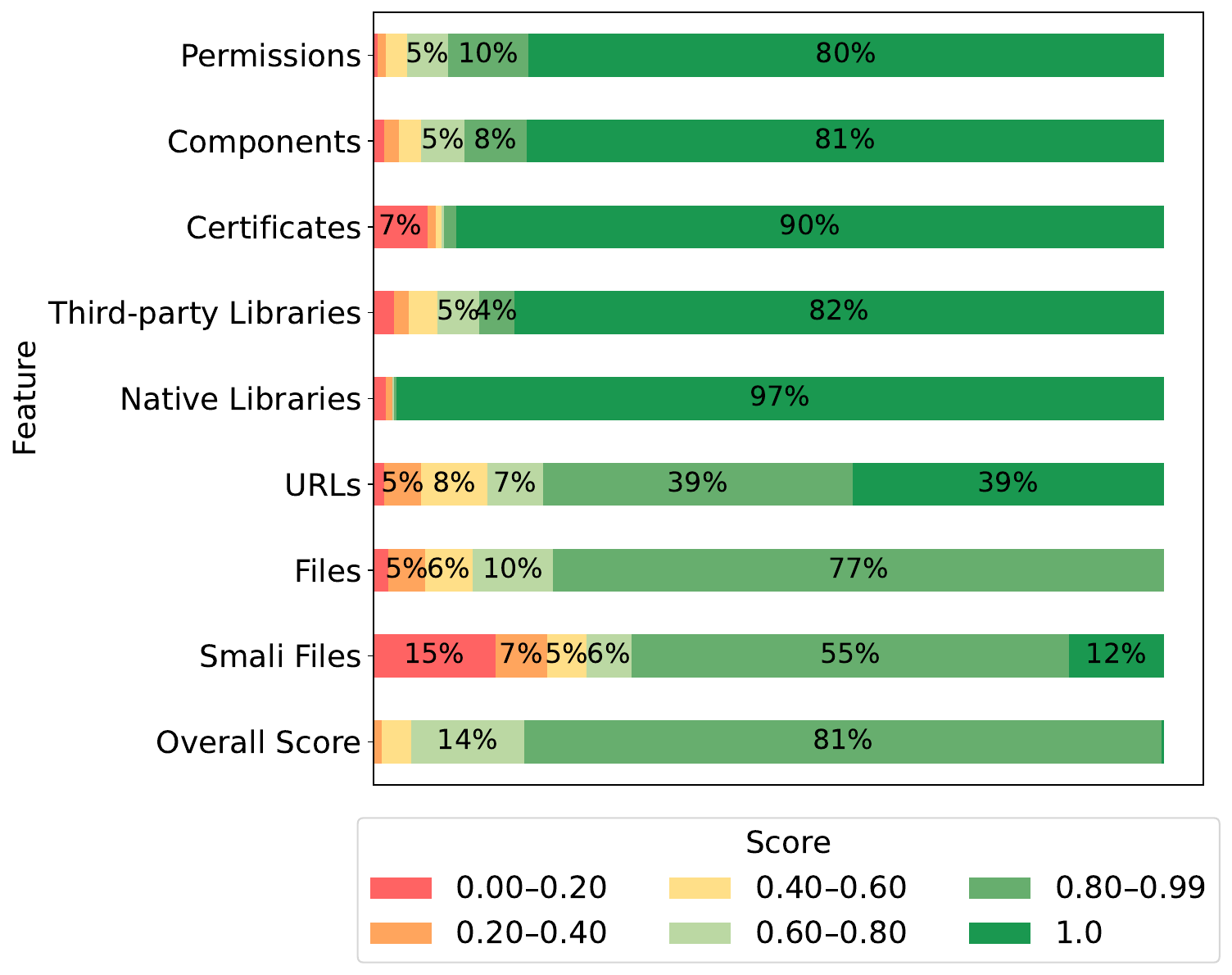}
    \caption{Distribution of similarity scores for different app features, categorized by score ranges. Values below 3\% are omitted for better visualization.}
    \label{fig:rq4bars}
\end{figure}

\noindent
\textbf{Results.}
We then compared the sampled GeoTwins using the custom similarity score tool introduced in Section~\ref{sec:analysis} (similarly to what we did in RQ3). Table~\ref{tab:rq4Results} reports the average, median, and correlation with the overall score for each feature. Figure~\ref{fig:rq4bars} further illustrates the distribution of the similarity score for each feature.

\begin{table}[h]
    \centering
    \caption{Similarity metrics between GeoTwins.}
    \label{tab:rq4Results}
    \begin{adjustbox}{width=0.6\linewidth}
\begin{tabular}{@{}l|c|c|c@{}}
\multicolumn{1}{c|}{\textbf{Feature}} & \textbf{Average} & \textbf{Median} & \textbf{Correlation with Overall} \\ \midrule
Permissions                             & 0.95 & 1.00 & 0.72 \\
Components                              & 0.94 & 1.00 & 0.80 \\
Certificates                            & 0.92 & 1.00 & 0.29 \\
Third-party Libraries                   & 0.93 & 1.00 & 0.78 \\
Native Libraries                        & 0.98 & 1.00 & 0.43 \\
URLs                                    & 0.87 & 0.97 & 0.75 \\
Files                                   & 0.87 & 0.97 & 0.78 \\
Smali Files                             & 0.76 & 0.98 & 0.83 \\ \midrule
\textbf{Overall Score}                  & \textbf{0.90} & \textbf{0.98} & \textbf{1.00}                                                      
\end{tabular}
    \end{adjustbox}
\end{table}

As it can be observed, GeoTwins are largely similar across most features: permissions, components, libraries, and certificates have high median scores (1.0), indicating full similarity in at least half of all samples. The largest divergences appear in smali code (avg. 0.76) and files (avg. 0.87), which also have the strongest correlation with the overall score, suggesting code-level changes are the primary source of divergence. 

In over 80\% of GeoTwins, permissions, components, certificates, and third-party libraries are identical. Even in features with more variability --like smali files, files, and URLs-- differences are relatively modest. Only 33\% of smali file comparisons fall below a similarity score of 0.8, and fewer than 25\% of file or URL comparisons do. No GeoTwins are completely identical overall (due in part to changes in package name and signing), but nearly 35\% have an overall similarity score above 0.99. This supports the hypothesis that GeoTwins are often minor variants of the same app, republished for regional targeting with limited underlying changes.

Our analysis indicates that most changes made to the APKs before republishing them in different regions are minimal. For instance, as seen in Figure~\ref{fig:guidatvIcons}, some applications are republished with different icons depending on the target country. These changes, while mostly cosmetic, affect the files similarity score, as the internal icon file is modified. More critically, we observed cases where republished apps differ in their declared permissions and even their components. This is particularly concerning because it raises questions about whether users in different countries are subject to different risks, despite apparently using the same application.

\noindent
\textbf{Illustrative Examples.} 
To further explore this topic, we closely examined three examples from our dataset, which exhibited differences across some of their features. In the case of the \textit{Wine} app --with a permissions score of \num{0.952}--, we found that the version with the package name \texttt{br.com.wine.app} declares two additional permissions not present in \texttt{br.com.wine.app.mx}: \path{ACCESS_ADSERVICES_TOPICS} and \texttt{ACCESS\_ADSERVICES\_CUSTOM\_AUDIENCE}. These extra permissions are related to advertising and their inclusion in only one APK raises questions about its underlying motivation. 

Meanwhile, in the case of the \textit{Knihovrátok} app (or \textit{Knihovrátek}, depending on the package name), which has a components score of \num{0.889}, we discovered that the version with the package name \texttt{sk.martinus.knihovratok} declares some additional components not found in \path{cz.martinus.knihovratek}. Specifically, it includes one extra service and two extra content providers: the service is \path{com.baseflow.geolocator.Geolocator.LocationService}, and the providers are \texttt{com.crazecoder.openfile.FileProvider} and \texttt{androidx.core.content.FileProvider}. The inclusion of the geolocation service is particularly concerning, as it may allow the Slovak version of the app to access users’ location data, a feature apparently absent in the Czech version. Again, these discrepancies raise concern about the motivations behind publishing regional variants.

Finally, we have the example shown in Figure~\ref{fig:unisonLeagueScreenshots}. The \textit{Unison League} GeoTwins have an overall similarity score of \num{0.891}, with notable differences across most features --especially in certificates and permissions--. It is uncommon to find different certificates among GeoTwins, but closer inspection reveals that the Japanese version is developed by a different team within the same company (\textit{Ateam}\footnote{\url{https://www.ateam-entertainment.com}}): the \textit{ito} team for Japan and the \textit{imai} team for the international version. Additionally, as noted in Section~\ref{sec:introduction}, the Japanese version requests the \texttt{ACCESS\_FINE\_LOCATION} permission, while the international version includes \texttt{MOUNT\_UNMOUNT\_FILESYSTEMS}. These differences suggest permissions are adapted to comply with region-specific regulations (e.g., GDPR~\cite{EUDataProtection}) rather than solely the game’s needs.

\begin{answer}
\textbf{Answer to RQ4:} \textit{GeoTwins are largely similar across most features. Our analysis shows that over 80\% of GeoTwin pairs share identical permissions, components, certificates, and third-party libraries. Differences mainly appear in smali code, URLs, and other files, though these are generally minor. Nearly 35\% of GeoTwins have a similarity score above 0.99, supporting the idea that these apps are mostly repackaged with small tweaks for regional use. Permission differences likely reflect local regulatory compliance rather than functional changes.}
\end{answer}
\section{Discussion}
\label{sec:discussion}

Our findings reveal that Android apps exhibit substantial, often hidden, regional variation at both the binary and behavioral levels. These results have implications for researchers, tool designers, developers, and regulators.

\noindent
\textbf{\bcircle{1} Impact on Research Methodologies.}
A recurring assumption in prior work on Android apps is that a single version of an app is representative of all its versions. Our results challenge this assumption: apps frequently exhibit region-specific behaviors, permissions, and libraries. As a result, research that relies on a single country-specific version—such as studies of permission evolution, or privacy leakage—may suffer from geographic bias. For example, a benign version of an app in one region might request significantly fewer permissions than its counterpart in another, potentially leading to misclassification or skewed conclusions.
Similarly, the reproducibility and replication of studies can be undermined when seemingly identical apps behave differently across regions. Silent, location-based variations in app binaries mean that reproducing prior results may inadvertently analyze a different regional variant, making conclusions non-replicable. Our dataset enables the research community to explicitly account for such regional differences.

\noindent
\textbf{\bcircle{2} Implications for Longitudinal Studies.}
Many longitudinal analyses track apps by package name, implicitly assuming that a package corresponds to a single product. The concept of \emph{GeoTwins} shows that this assumption can be misleading: the same brand may be represented by multiple package names across different countries. These findings suggest that future analyses should consider grouping apps into \emph{GeoFamilies}, enabling a more comprehensive understanding of cross-country variation in privacy, advertising, and functionality.

\noindent
\textbf{\bcircle{3} Insights for Developers.}
Our findings suggest that regional divergence is not always security-driven. Cultural differences also influence app composition: for example, we found that Japanese apps rarely embed the \texttt{com.facebook} library. Developers seeking to expand internationally can use our dataset as a resource to better understand country-specific norms and tailor app features accordingly.

\noindent
\textbf{\bcircle{4} Consequences for Users and Regulators.}
GeoTwins raise important questions about transparency and informed consent. Users downloading visually identical apps from different regions may unknowingly receive versions with more invasive permissions, less robust privacy safeguards, or different embedded services. Current app store interfaces provide little to no disclosure about such variations, despite their potential impact on privacy and security. Regulators could consider requiring platforms to disclose regional customizations more explicitly to ensure consistent user protections and to enforce compliance with local regulations.

Overall, our work highlights the need to rethink how research, regulation, and development practices account for regional app differences, and provides the first large-scale dataset to support such efforts.
\section{Limitations}
\label{sec:limitations}

In this section, we acknowledge the potential limitations of our study, explain how we mitigated them, and suggest how these challenges can be addressed in future research.

\noindent
\textbf{Geographical Coverage.}
The list of countries included in our study is not exhaustive and may not fully represent the global diversity of app ecosystems. To mitigate this limitation, we selected countries from different macro-regions (e.g., Asia, North America, Europe, and Africa) to provide a broad, albeit non-comprehensive, geographic perspective.

\noindent
\textbf{Static Analysis Only.}
Our methodology relies solely on static analysis due to the large volume of apps examined. While static techniques offer scalability and reproducibility, they cannot capture runtime behaviors such as dynamic permission requests or API calls. Future work may incorporate dynamic analysis tools --such as Monkey or other automated UI exercisers-- to uncover further region-specific behaviors that manifest only during execution.

\noindent
\textbf{GeoTwins Dataset.}
The strict filtering applied in the automated pipeline developed for constructing the GeoTwins dataset may lead to the exclusion of some potential GeoTwins that do not fully meet our predefined criteria. However, the GeoTwins dataset represents a novel contribution, and in order to provide a reliable resource for future researchers, we deliberately chose a highly conservative approach. This prioritizes precision and minimizes false positives, even at the cost of potentially overlooking some legitimate GeoTwins. 
\section{Related Work}
\label{sec:relatedwork}
In this section, we review prior research on geographic variation in mobile applications.

Kumar et al.~\cite{kumar2022large} conducted a large-scale study of Android apps distributed across 26 countries, analyzing approximately \num{5000} apps. Their work mainly focused on \textit{GeoBlocking}, i.e., when apps are unavailable in certain countries, and the motivations behind this phenomenon, such as government-requested takedowns or developer blocking. They also showed that apps with the same package name can vary in permissions and privacy policies depending on the region from which they are downloaded. More recently, Guo et al.~\cite{guo2025code} introduced FREELENS, a framework for identifying geographic feature differences (GFDs) at the code level. By analyzing over \num{21000} apps across ten countries, they uncovered significant regional disparities in areas such as advertising, authentication, and data handling. However, their study exclusively examined apps with the same package name and explicitly acknowledged the absence of \textit{GeoTwins} in their analysis. For example, in their limitations section, they referenced the GeoTwin pair \texttt{com.americanexpress.android.acctsvcs.uk} and \texttt{com.americanexpress.android.acctsvcs.us}, but did not analyze it, whereas our analysis includes this pair. In addition to covering an unexplored aspect like \textit{GeoTwins} --apps with similar branding but different package names for targeted regions (see RQ4)-- we also addressed another limitation common to previous studies. Indeed, we took into account the Android App Bundle format by analyzing complete app bundles and isolating the \texttt{base.apk}, thus avoiding expected differences that could be present in the split files (see RQ3). 

Various sources have also highlighted the cultural and economic dimensions of regional differences in mobile app ecosystems, discussing how app markets reflect local preferences, platform policies, and socio-economic conditions~\cite{li2021finding, peltonen2018hidden, algoace2024cultural, kravtsov2019region, wang2024global}. Our findings contribute to this discussion by providing empirical evidence of regional differentiation in mobile apps. We observed distinct preferences in third-party libraries (e.g., particularly in Japanese apps), and the prevalence of country-exclusive apps suggests that cultural and economic factors play a significant role in shaping mobile ecosystems. Future work could build on our analysis by incorporating user reviews and ratings to better understand regional user expectations and sentiments.

\section{Conclusion}
\label{sec:conclusion}

This paper investigated how Android applications vary geographically, moving beyond surface-level accessibility to uncover deeper compositional differences. We introduced and systematically explored two previously unexplored phenomena: \textit{GeoTwins} --functionally similar apps distributed under distinct package names to target different regions-- and surprising regional variability within supposedly consistent \texttt{base.apk} files (e.g., different permissions and third-party libraries).

Our analysis revealed that these \textit{GeoTwins}, despite shared branding, can diverge significantly in critical aspects like the smali files (e.g., the code of the app), files, and URLs. Furthermore, the discovery of regional variations in \texttt{base.apk} files challenges the assumption of a consistent core for apps across locations, indicating hidden customizations. These findings demonstrate that location-based differentiation is more profound than previously understood, impacting fundamental app structure and behavior.

Our study highlights that geographical location shapes not just app availability, but the very nature of the software users interact with, raising important considerations for privacy equity, platform transparency, and user experience in the global app ecosystem. Future work could extend this comprehensive study to \textit{GeoFamilies} — collections of multiple regional variants of the same app — and could also incorporate an analysis of user reviews and ratings to better understand regional user expectations and sentiments.

\bibliographystyle{ACM-Reference-Format}
\bibliography{references}

\end{document}